\documentclass[12pt]{iopart}
\usepackage{graphicx}

\def\simleq{\; \raise0.3ex\hbox{$<$\kern-0.75em \raise-1.1ex\hbox{$\sim$}}\; }
\def\simgeq{\; \raise0.3ex\hbox{$>$\kern-0.75em \raise-1.1ex\hbox{$\sim$}}\; }
\newcommand{\eV}{{\rm eV}}
\newcommand{\keV}{{\rm keV}}

\newcommand{\GeV}{{\rm GeV}}
\newcommand{\TeV}{{\rm TeV}}
\newcommand{\PeV}{{\rm PeV}}

\newcommand{\kpc}{{\rm kpc}}
\newcommand{\pc}{{\rm pc}}
\newcommand{\cm}{{\rm cm}}
\newcommand{\km}{{\rm km}}
\newcommand{\muG}{\mu{\rm G}}

\newcommand{\s}{{\rm s}}
\newcommand{\yr}{{\rm yr}}
\newcommand{\sr}{{\rm sr}}

\begin{document}
\title{Diffuse Neutrino and Gamma-ray Emissions of the Galaxy above the TeV}

\author{Carmelo Evoli$^1$, Dario Grasso$^2$, Luca Maccione$^{1,3}$}

\address{$^1$ SISSA, via Beirut, 2-4, I-34014 Trieste}
\address{$^2$ INFN, Sezione di Pisa, Largo Bruno Pontecorvo, 3, I-56127 Pisa}
\address{$^3$ INFN, Sezione di Trieste, Via Valerio, 2, I-34127 Trieste}

\eads{\mailto{evoli@sissa.it}, \mailto{dario.grasso@pi.infn.it}, \mailto{maccione@sissa.it}}

\begin{abstract} 
We simulate  the  neutrino and $\gamma$-ray emissions of the Galaxy which are originated from the hadronic scattering of cosmic rays (CR) with the interstellar medium (ISM). 
Rather than assuming a uniform CR density, we estimate the spatial distribution of CR nuclei by means of numerical simulations. We  consider several models of the galactic magnetic field and of the ISM distribution finding only a weak dependence of our results on their choice. We found that by extrapolating the predicted $\gamma$-ray spectra down to few GeV  we get a good agreement with EGRET measurements. Then, we can reliably compare our predictions with available observations above the TeV both for the $\gamma$-rays and the neutrinos. 
We confirm that the excesses observed by MILAGRO in the Cygnus region and by HESS in the Galactic Centre Ridge
cannot be explained without invoking  significant  CR over-densities in those regions.   
Finally, we discuss the perspectives that a km$^3$ neutrino telescope based in the North hemisphere has to measure the diffuse emission from the inner Galaxy. 
\end{abstract}

\section{Introduction}

It is well known that the interaction of  cosmic ray (CR) nuclei with the interstellar medium (ISM) should give rise to diffuse $\gamma$-ray and neutrino emissions \cite{Berezinsky:book,Gaisser:book}. Several satellites (see \cite{Bloemen:89} for a review), especially EGRET \cite{Hunter:97,Cillis:05}, already observed the $\gamma$-ray diffuse emission of the Galaxy up to $\sim 10~\GeV$.
Around the GeV, the hadronic origin of a large fraction of that emission is testified by its good correlation with the distribution of the interstellar hydrogen. GLAST observatory \cite{Glast} should soon provide a deeper insight into the nature of that emission as it will probe the $\gamma$-ray sky with much better sensitivity and angular resolution than EGRET and extend the explored energy window up to 300 GeV.
Above that energy, however, satellite observatories can hardly provide significant data.

Ground based experiments are divided in two  main  groups: extensive air shower arrays and atmospheric Cherenkov telescopes. In the former class,  experiments like MILAGRO  \cite{Milagro} and TIBET \cite{Tibet}  can probe large 
regions of the  sky at energies larger than few TeV's   with an angular resolution of  several degrees. Those experiments 
already provided   interesting constraints on the diffuse emission from the galactic disk \cite{Milagro:obs, Tibet:obs} 
and, in  the case of MILAGRO,  a  diffuse emission from the Cygnus region has been recently observed \cite{Abdo:2006}.
Unfortunately, none of those experiments covers a sky window encompassing the Galactic Centre (GC).
Atmospheric Cherenkov telescopes, like HESS \cite{HESS} and MAGIC \cite{MAGIC}, are sensitive to $\gamma$-rays  with energy in the range $0.1 \simleq E \simleq 100~\TeV$ and have a very good angular resolution (better than $0.1^\circ$).  However, although these instruments can search for a diffuse emission in some limited regions of the sky, they are mainly dedicated to observe quite localised sources.

Neutrino telescopes (NTs)  may  soon provide a new valuable probe of high energy CR physics which can be  complementary to $\gamma$-ray observatories. The large detection volume, large field of view, available observation time, and relatively good angular resolution (which should be better than $1^\circ$ for water Cherenkov NTs looking for up-going muon neutrinos) make, at least in principle,  these instruments well suited to search for the diffuse neutrino emission of the Galaxy.
\\The AMANDA telescope, which is operating at the South Pole, put already an upper limit on the $\nu_\mu$'s flux from the GP \cite{Kelley:05} (see \sref{sec:neutrinos}) and ICECUBE \cite{Icecube} should significantly improve that limit.

However, in this context the most promising NTs are those based in the North hemisphere (ANTARES \cite{Antares},  NEMO \cite{NEMO} and NESTOR \cite{NESTOR} and especially their planned $\km^3$ upgrade, the Km3NeT project \cite{km3net}) as they will be sensitive to up-going $\nu_\mu$'s coming from the GC region. Furthermore, since these instruments will be deployed in water, their angular resolution should be better than NT at the South pole improving their chances to disentangle the signal from the background.

From the theoretical point of view, several calculations of the $\gamma$-ray and neutrino diffuse fluxes above the TeV  have been already performed. In \cite{Stecker:79, Berezinsky:93,Ingelman:96} the CR distribution in the Galactic disk 
was assumed to be homogeneous so that the secondary $\nu$ and $\gamma$-ray fluxes come out to be
 proportional to the gas column density  (in \cite{Stecker:79,Ingelman:96} a uniform gas distribution was assumed). 
 While that was a very reasonable approximation to start with, a more accurate analysis should account for the  
 inhomogeneous CR  distribution which arises due to the not uniform scatter of  sources.  
It should be noted that, since CR sources are likely to be related to supernova remnants (SNR) and those objects are  most abundant in the gas (target) rich regions, the secondary $\gamma$-ray and neutrino emissions may be significantly enhanced with respect to the case in which a uniform CR distribution is assumed. The CR homogeneity assumption 
was released in more recent works where the high energy nuclei distribution  was modelled by solving the spatial diffusion equation \cite{Strong:98,Strong:04}. Those papers, however, were mainly addressed to model the diffuse $\gamma$-ray emission below the TeV and did not consider neutrinos.
In \cite{Candia:05} a different, and promising, approach was considered to solve the diffusion equation and to model the neutrino emission up to the PeV. In that paper, however, several approximations were done which did not allow to reach an angular accuracy better than $10^\circ$.  

The aim of this work is to model both the  diffuse $\gamma$-ray and neutrino emissions of the Galaxy above the TeV with a better  accuracy. That is necessary in order to be able to establish the detection perspectives of air shower arrays and neutrino telescopes and to interpret correctly forthcoming observations. Our approach is comprehensive as we carefully  account both for the  spatial distribution of CR nuclei and for that of the ISM.  

In \sref{sec:galaxy} we  start by discussing the main properties of the ISM.
We  pay special attention to the distribution of SNR (see \sref{subsec:snr}) and to that of the  atomic and 
molecular hydrogen (\sref{subsec:gas}). In both cases we adopt models which we apply for the first time in this 
context.  Details of gas models and their comparison with those adopted in 
previous works are given in  the  \ref{sec:appendix_gas}. 
In \sref{sec:diffusion} we model the distribution of the CR nuclei in the Galaxy by solving numerically 
the diffusion equation. Our approach is similar to that sketched in \cite{Candia:05}  that we fully 
exploit here.  Differently from \cite{Strong:98,Strong:04} we use expressions for the diffusion coefficients as determined from MonteCarlo simulations of charged particle propagation in turbulent magnetic fields  \cite{Casse:02,Candia:02}.
This allows us to get  more detailed CR distributions above the TeV  and to  test  how much the large uncertainties in the knowledge of the turbulent component of the GMF 
(see \sref{subsec:mf}) affect our predictions. 
In \sref{sec:neutrinos} we combine the simulated CR distributions and the gas models to 
map the expected $\nu_\mu$  and $\gamma$-ray emissions. 
Then, in \sref{sec:experiments} we  compare our predictions with available experimental results. 
In \sref{sec:km3} we  briefly discuss the perspectives that a ${\rm km}^3$ NT 
 to be built  in the  Mediterranean sea has to detect the muon neutrino emission from the inner 
 Galaxy. Finally, in \sref{sec:conclusions} we summarise our conclusions.

\section{The spatial structure of the ISM} \label{sec:galaxy}

In order to assess the problem of the propagation of CRs and their interaction with the  ISM we need the knowledge of three basic physical inputs, namely:
\begin{enumerate}
 \item the distribution of SuperNova Remnants (SNR) which we assume to trace that of  CR sources;
 \item the properties of the  Galactic Magnetic Field (GMF) in which the propagation occurs;
 \item the distribution of the diffuse gas providing the target for the production of $\gamma$-rays and
neutrinos through hadronic interactions.
\end{enumerate}

Since all these inputs are affected by large uncertainties and systematics, we need to estimate how their poor knowledge 
is transferred in the resulting $\gamma$-ray and $\nu$ emission. Therefore, we compare the results obtained using 
several models.  
In the following we assume cylindrical symmetry  and adopt the Sun galactocentric distance $r_\odot = 8.5~\kpc$.

\subsection{The SNR distribution in Galaxy}\label{subsec:snr}

The rate of galactic SN explosions, as inferred from observations of SN in external galaxies similar to the Milky Way \cite{Cappellaro:97}, is roughly
\begin{equation}
R_{\rm I} \sim \frac{1}{250} \;\yr^{-1} \qquad R_{\rm II} \sim \frac{1}{60}\; \yr^{-1}\ ,
\label{I-II-rates}
\end{equation}
respectively for type I and type II SN. The total rate is $\simeq 1/48~\yr^{-1}$ which is in reasonably good agreement with records of historical SN. Since we will normalise the CR injection rate by requiring the simulated CR flux to meet the observed value at Earth position, we are only interested in the $R_{\rm I}$ and $R_{\rm II}$ ratio.

What is even more crucial to our analysis is the {\it spatial distribution} of type I and II SN.
Since type-Ia and core-collapsed SN originate from different stellar populations their spatial distribution, as well as their rate, are different. 

Several methods to determined the  SNR  distribution in the Galaxy  are discussed in the literature. 
One of the most commonly adopted methods  is that to estimate the SNR distances on the basis of the surface brightness - distance ($\Sigma - D$) relation  \cite{Case:1998qg}. 
Such an analysis does not cover the GC region.  Furthermore, several doubts have been risen on
 the accuracy of that method  as it is plagued by  a  number of systematics  concerning the  
 completeness of the available SNR catalogue and the proper handling  of  selection effects 
 \cite{Green:96,Green:05}. Here we adopt a SNR distribution a distribution which is inferred  from observations
  of related objects, such as pulsars or progenitor stars, as done e.g. in \cite{Ferriere} which, in our opinion, 
  is a safer approach. 

In \cite{Ferriere} the spatial distribution of type-Ia SNR adopted  was assumed to follow that of old disk stars which have an exponential scale length $\simeq 4.5~\kpc$ along $r$ and an exponential scale height $\simeq 300~\pc$. Therefore
\begin{equation}
{\cal R}_{\rm I} (r,z) =   K_{\rm I} \exp \left( - \frac{r - r_{\odot}}{4.5~{\rm kpc}} \ 
- \frac{|z|}{0.3~{\rm kpc}} \right)~,
\label{SNR1}
\end{equation}
where $K_{\rm I}$ is a normalisation factor. 
Although type-Ia SN are globally less frequent than core-collapsed SN, their rate is dominating in the inner few kpc's of the Galaxy.

To trace  core-collapsed originated SNR one may use either HII regions, which are produced by their 
luminous progenitors, or pulsars, which are a likely left-over of the collapse. 
The pulsar distribution at birth was estimated to be \cite{Ferriere,Narayan:87,Narayan:90}  
\begin{equation}
{\cal R}_{\rm II}(r,z) \, =  K_{\rm II}~ f(z)\  \left\{   
\begin{array}{ll}
3.55 \ \exp \left[ - \left( { \displaystyle r - 3.7 \ {\rm kpc} 
                              \over
                              \displaystyle 2.1 \ {\rm kpc} } 
                     \right)^2 \right] \ , 
& r < 3.7 \ {\rm kpc}
\nonumber  \\
\noalign{\medskip}
\exp \left[ - { \displaystyle r^2 - r_\odot^2 
                \over 
                \displaystyle (6.8 \ {\rm kpc})^2 } \right] \ , 
& r > 3.7 \ {\rm kpc}
\end{array}
\right.
\label{SNR2}
\end{equation}
where
\begin{equation}
f(z) = 
0.79 \ \exp \left[ - \left( {z \over 0.2 \ {\kpc}} \right)^2 \right] 
+ 0.21 \ \exp \left[ - \left( {z \over 0.6 \ {\kpc}} \right)^2 \right]~.
\end{equation}
While the absolute values of $ K_{\rm I}$ and $ K_{\rm II}$ are irrelevant here, their ratio is needed  to normalise the relative weights of type-I and type-II SNR distributions.
By requiring that  $R_{\rm II}/R_{\rm I} = 4.2$,  as it follows from (\ref{I-II-rates}), we find $K_{\rm II}/K_{\rm I} \simeq 7.3$.

\begin{figure}[!th]
 \centering
 \includegraphics[scale=0.5]{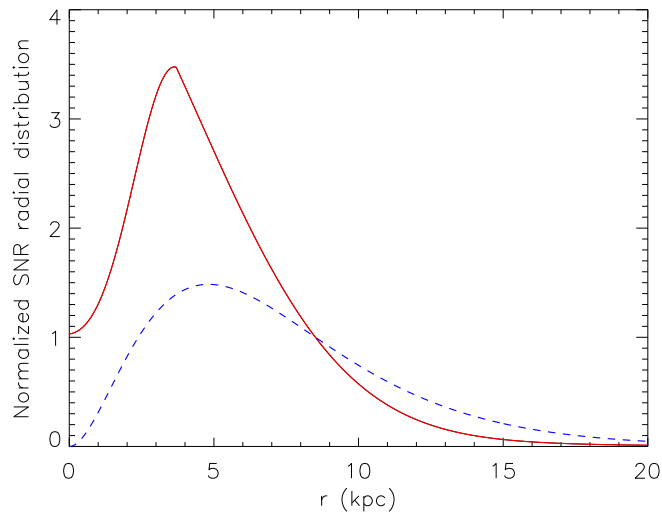}
\caption{The SNR radial distribution $f(r, z=0)$ is shown in arbitrary units versus the distance from 
the GC. The distribution is such that, in cylindrical symmetry, $2\pi\int rf(r,z) \,dr\, dz$ represents the total SN rate in the whole galaxy. The upper line (red, continue) is that derived in \cite{Ferriere} which we adopt here, while the lower one (blue, dashed) is from \cite{Case:1998qg}. They are both normalised to unit at $r = r_\odot = 8.5~\kpc $.}  \label{fig:SNR}
\end{figure}

In \fref{fig:SNR} we show the total SNR radial distribution as obtained in that way and compare it with that given in \cite{Case:1998qg}.
Both distributions have been normalised so to take the same value at the solar circle where observations are most reliable.
It is evident that in the inner Galaxy the SNR rate distribution  which we adopt in this work, is significantly higher than that given in \cite{Case:1998qg}.
Rather, we have a relative good agreement with the distribution adopted in \cite{Lorimer:04}
(see also  \cite{Strong:04b}) which is based on pulsars, and with independent observations of the 
$1809~\keV$ line of $^{26}$Al which is thought to be a reliable tracer of SNs \cite{Knodlseder:1999gk,Plueschke:2001dc}. 

We assume the spatial distribution of the CR sources to coincide with that of SNRs and the CR energy spectrum at source to 
be everywhere a power-law with exponent $\beta$ ($\beta$, which refers to the injection spectrum, should not be
confused with the slope of the observed CR spectrum). Therefore, the source term in our diffusion equation will be
\begin{equation}
\label{eq:source_term}
Q(E,r,z) = q(r,z)~E^{-\beta} = K\left( {\cal R}_{\rm I}(r,z) +  {\cal R}_{\rm II}(r,z)\right)~E^{-\beta} \;.
\end{equation}
Its absolute normalisation $K$ and spectral slope $\beta$ will be fixed so to reproduce the observed spectrum at Earth below the knee.
For protons \cite{Horandel:03} 
\begin{equation}
\label{p_spectrum}
I_p(E)  \simeq 8.7  \times 10^{-6}~\left(\frac {E}{1~\TeV}\right)^{-2.7}   
(\TeV~\cm^2~\s~\sr)^{-1}\;.
\end{equation}

\subsection{Regular and random magnetic fields}\label{subsec:mf}

The Milky Way, as well as other spiral galaxies, is known to be permeated by large-scale,
so called {\it regular}, magnetic fields.  The orientation and strength of those fields is 
measured mainly by means of Faraday Rotation Measurements (RMs) of polarised radio sources.  From those 
observations it is known that  the 
regular field in the disk of the Galaxy  is prevalently oriented along the disk plane and it seems to follow the galactic arms as observed in other spiral galaxies. 
 According to \cite{Han:06},  its strength at the Sun position is $B_o \equiv B^{\rm disk}_{\rm reg}(r_\odot, 0)  = 2.1 \pm 0.3~\muG$
while at smaller radii  $\displaystyle
B^{\rm disk}_{\rm reg}(r) = B_0 \exp \left\{ - \frac{r - r_\odot}{r_B} \right\}$
where $r_B = 8.5 \pm 4.7~\kpc$.   A  $1/r$ profile seems to give a worst fit of data. 
Unfortunately, observations are not significant  for $r < 3~\kpc$.
Most likely \cite{Han:06} the regular field in the disk has a 
bi-symmetric structure (BBS)  with a counterclockwise field in the spiral arms and clockwise in the interarm regions. Concerning its vertical behaviour, it is generally assumed that ${\bf B}^{\rm disk}_{\rm reg}$  decreases exponentially for increasing values of $|z|$ with a scale height of few hundred parsecs.  
There are increasing evidences   that the field is symmetric  for 
$z \rightarrow -z$ (BBS-S) \cite{Beck:01}.

Superimposed to the regular field a random, or turbulent,  component of the GMF is known to
 be present.  In the disk,  this component is comparable to, or even larger than, the regular one. Indeed, the locally  observed  rms value of the total field is about $6 \pm 2~\muG$.
which is $2-4$ times larger than $B_{\rm reg}(r_\odot, 0)$. 
From polarimetric measurements  of stellar light and RMs of close pulsars  it has been
inferred that the GMF is chaotic on all scales below $L_{\rm max} \sim 100~\pc$. 
The power spectrum of the GMF fluctuations is poorly known. 
Observational data, obtained from RM of pairs of close pulsars,  
are compatible  with a Kolmogorov spectrum, i.e.  
$B^2(k) \propto k^{-5/3}$,  though with a very large uncertainty (see e.g. \cite{HanFer:04} and ref.s therein). 
    
What is most relevant here is the  galactic magnetic  halo (MH) since  most of CR propagation takes place
outside the disk.    The  regular component of the MH has been studied  by means of RMs of polarised 
 extra-galactic radio sources.  Those observations showed that  ${\bf B}^{\rm halo}_{\rm reg}$ 
 is mainly azimutal,  the same  as ${\bf B}^{\rm disk}_{\rm reg}$,  and that its vertical scale  
 height is $z_r \simeq 1.5~\kpc$  \cite{Han:94}.
The radial  behaviour of  ${\bf B}^{\rm halo}_{\rm reg}$ is  poorly known and  
it is generally assumed that it traces that of $B^{\rm disk}_{\rm reg}$.  Here we share the 
same attitude. It is worth noticing  that  the vertical symmetry of the MH 
might be opposite with respect to that in the disk: this fact may have relevant consequences
for the propagation of CRs with  $E/Z \simgeq 10^{15}~\eV$ \cite{Candia:02}. 

In the following we assume  a symmetric structure so that the regular component of the MH can
 be considered as an extension  of  ${\bf B}^{\rm disk}_{\rm reg}$ and we can combine both in
 a unique simple structure:
\begin{equation}
\label{Breg}
B_{\rm reg}(r,z) = B_0 \exp \left\{ - \frac{r - r_\odot}{r_B}\right\} \; 
\frac{1}{2\cosh(z/z_r)}~,
\end{equation}
with $z_r = 1.5~\kpc$ and $r_B$ left as a free parameter.
Similarly to  \cite{Candia:02} we use a $\cosh$ function
to regularise the exponential vertical profile of the regular and turbulent 
components of the GMF at $z = 0$. 

Concerning the properties of the random MF component in the halo,
 very little is known from observations.   
Both the vertical extension of the radio halo and the isotopic relative abundances of CR
suggest that the vertical scale height of the random fields is significantly larger than 
that of the regular one, $z_t \simeq 3 - 5\;\kpc$ or  larger. 
It is also likely that the turbulence strength increases in the regions with the 
highest star-forming activity.  We account for this possible radial dependence 
of $B_{\rm turb}$ by means of the parameter 
$\displaystyle \sigma(r) \equiv \frac{\langle B_{\rm ran} \rangle_{\rm rms}(r,0)}
{B_{\rm reg}(r,0)}$.
Therefore, we write 
\begin{equation}
\label{Bran}
B_{\rm ran}(r,z) = \sigma(r) \;B_{\rm reg}(r,0) \;\frac{1}{2\cosh(z/z_t)}\; .
\end{equation}
Due to the relative small number of extra-galactic radio sources,  the maximal scale 
of magnetic fluctuations $L^{\rm halo}_{\rm max}$  and their power spectrum  in  the MH are practically unknown. 
In the following we assume that $L^{\rm halo}_{\rm max}$ is the same as in the disk and 
consider only  Kolmogorov  and Kraichnan $\left( B^2(k) \propto k^{-3/2}\right)$
power spectra.

\subsection{The gas distribution}\label{subsec:gas}

The diffuse gas accounts for about 10-15\% of the total mass of the Galactic disk (not including dark matter) and its chemical composition is dominated by hydrogen (about $90.8~\%$ by number and $70.4~\%$ by mass) and helium ($9.1~\%$ by number and $28.1~\% $ by mass). Hydrogen is shared in three main components \cite{Ferriere}: ionised (HII, total mass $M_{\rm HII} \simeq  1 \times 10^9~M_\odot$), atomic (HI, $M_{\rm HI} \simeq  6 \times 10^9~M_\odot$) and molecular (${\rm H}_2$, $M_{{\rm H}_2} \simeq  1 - 2  \times 10^9~M_\odot$).
While all these components have comparable masses, their spatial distributions are quite different.
The much hotter HII has a scale height along the vertical axis considerably larger than the other hydrogen components ($h_{\rm HII} \simeq 1~\kpc$).  Therefore its contribution to the $\gamma$-ray and neutrino emissivity from the galactic plane is subdominant and can be neglected in the following.

The construction of galacto-centric radial density profiles for the HI and ${\rm H}_2$ intrinsically needs model dependent assumptions (see \ref{sec:appendix_gas} for further details). Therefore large uncertainties are involved in this operation. Here we consider two models, which we call gas models A and B, both for the atomic and molecular hydrogen, that will be discussed in details in \ref{sec:appendix_gas}.

\begin{description}
\item[Model A] It has been developed by Nakanishi and Sofue (NS) in \cite{Nakanishi:03} for the HI and by the same authors in \cite{Nakanishi:06} for the  ${\rm H}_2$.
\item[Model B] We construct model B by suitably combining the results of different analyses which have been separately performed for the disk and the galactic bulge. For the ${\rm H}_2$ and HI distributions in the bulge we use a detailed 3D model recently developed by Ferriere et al. \cite{Ferriere:07} on the basis of several observations. For the molecular hydrogen in the disk we use the well known Bronfman's et al. model \cite{Bronfman:88}. For the HI distribution in the disk, we adopt Wolfire et al. \cite{Wolfire:03} 2-dimensional model. Although  Ferriere et al. model is not cylindrically symmetric, we verified that by fitting their 3D distribution with a cylindrically  symmetric one and assuming a Gaussian vertical profile peaked on the GP, we get $\gamma$-ray and neutrino fluxes that, when integrated over windows larger than 1 degrees squared, differ very little from those obtained using the complete 3D models. For this reason, in the following we work only with averaged 2D distributions.
\end{description}

\begin{figure}[!th]
 \centering
 \includegraphics[scale=0.72]{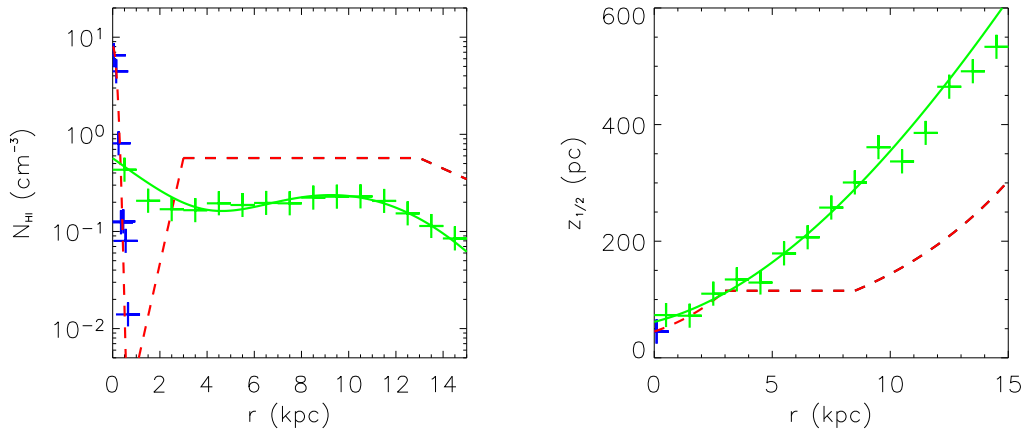}
 \caption{ Left panel: the ${{\rm HI}}$ volume density along the Galactic Plane ($z = 0$) is plotted versus 
 the radius $r$.  Green crosses display the binned distribution as given in \cite{Nakanishi:03} 
 while  and the continuos  line (green) is our  fit of those points (model A) .  Blue crosses are derived from
 \cite{Ferriere:07}. The dashed (red) line is our fit of those points with the distribution given in \cite{Wolfire:03}. 
  for $r > 3~ kpc$ (model B).   Right panel:  the  radial profile of the HWHM scale height ($z_{1/2}$) for the same models.}
 \label{fig:HIfit}
\end{figure}

\begin{figure}[!th]
 \centering
 \includegraphics[scale=0.72]{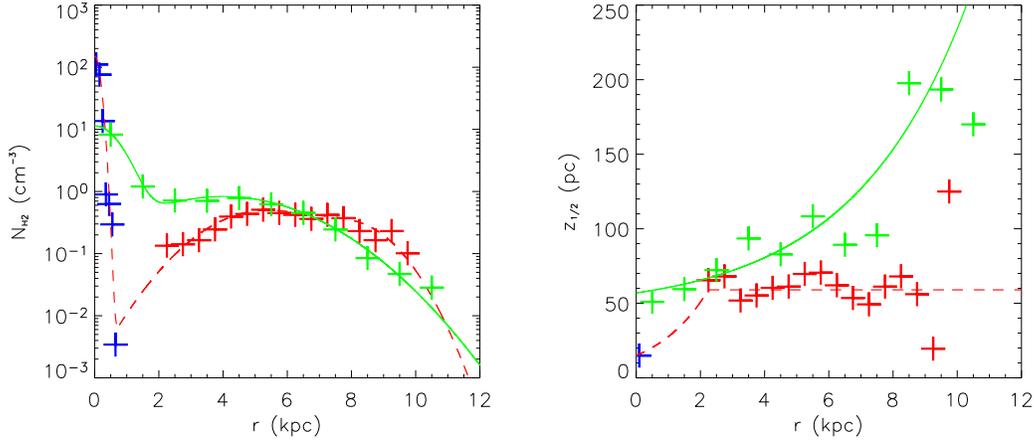}
 \caption{Left panel: the ${{\rm H}_2}$ volume density along the Galactic Plane as a function of $r$. 
 Green crosses display the binned distribution as given in \cite{Nakanishi:06}  and the continuos (green)  line 
 is our continuos fit of those points (model A) .  Red and blue  crosses are derived   (see text)  from  
 \cite{Bronfman:88} and \cite{Ferriere:07} respectively.  The  dashed (red) line is our combined fit
 of all those points (model B).
 Right panel: the radial profile of the HWHM scale height ($z_{1/2}$) for the same models.}
 \label{fig:H2fit}
\end{figure}

In \fref{fig:HIfit} and \fref{fig:H2fit} we show the volume density radial and vertical profiles of the HI and the ${\rm H}_2$ obtained with models A and B together with the continuos fits we use in our analysis without reporting observational errors as they are typically much smaller than systematic ones (the difference between model A and B will provide a glimpse of the amount of those uncertainties).

\begin{figure}[!th]
 \centering
 \includegraphics[scale=0.78]{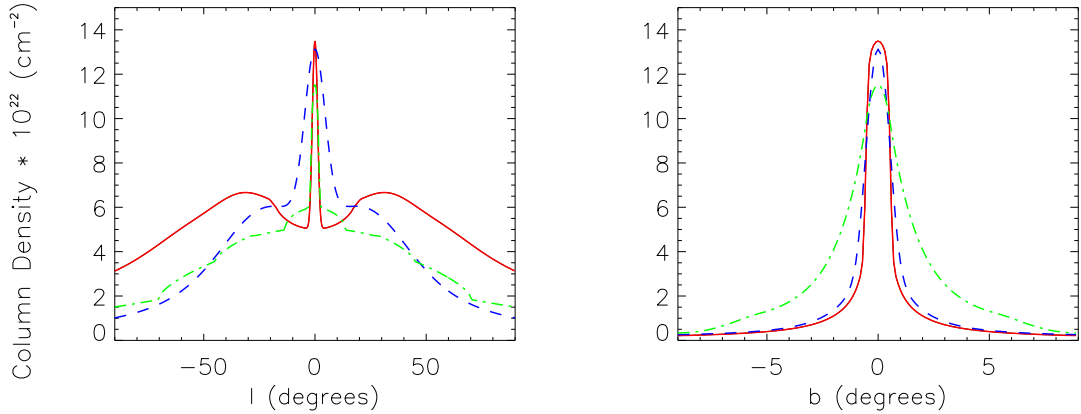}
\caption{Profiles of the hydrogen nuclei (HI + 2H$_2$) column density are shown along the galactic plane (left panel) and along the $l = 0$ line (right panel). The three diagrams correspond to model A (dashed, blue line), model B (continuos,  red line) and to the hydrogen distribution adopted in \cite{Berezinsky:93}.
Column densities are averaged over $1^\circ \times 1^\circ$ angular bins.} 
 \label{fig:cd_comp}
\end{figure}

While the gas densities in model A and B differ relatively little close to the solar circle (what is most relevant here is the ${\rm H}_2$) the main discrepancies arise in the galactic bulge. Indeed, that is the region where the uncertainties on the gas velocity are the largest. That discrepancy, however, has little consequences on the gas column density (see \fref{fig:cd_comp}) as it should since this quantity is almost directly related to the observed CO emission.

In the following we will use model B as our reference model since, in the central region of the GE, it provides a better fit of the $^{12}$CO emission survey \cite{Dame:01}. In \fref{fig:cd_comp} we also compare the hydrogen column density distributions obtained with models A and B with that used in \cite{Berezinsky:93} finding that the former are more narrowly peaked along the GP.
In the following we will assume that  helium is distributed in the same way as all hydrogen nuclei.

\section{Numerical Simulation of  CR diffusion in the Galaxy} \label{sec:diffusion}

The ISM is a quite turbulent magneto-hydro-dynamic (MHD) environment. Since the Larmor radius of high energy nuclei,
\begin{equation}
r_L(E)  = \frac{E}{Z e B_{\rm reg}} \simeq
0.1\  \left(\frac{E}{10^{2}~\TeV}\right) \ \left(\frac{B_{\rm reg}}{1~\muG}\right) \ {\rm pc}~,
\label{rL}
\end{equation}
is much smaller than the maximal scale of  the magnetic field fluctuations 
$L_{\rm max} \sim 100~\pc$, the propagation of those particles takes place in the 
spatial diffusion regime. 
The diffusion equation which describes such a propagation is (see e.g. \cite{Ptuskin:93}) 
\begin{equation}\label{j}
\nabla_i  J_i (E,r,z) \equiv - \nabla_i\left( D_{ij} (r,z) \nabla_j N(E,r,z)
\right)  = Q (E,r,z) 
\end{equation}
where $N(E,r,z)$ is the differential CR density averaged over a scale larger than $L_{\rm max}$, $Q (E,r,z) $ is the CR source term (\ref{eq:source_term}) and $D_{ij} (E,r,z)$ are the spatial components of the diffusion tensor.
We assume cylindrical symmetry and use Einstein notation for repeated indexes.
\Eref{j} does not contain energy loss/gain terms.
This is justified by the fact that in the high energy range relevant for our problem 
($E_{\nu~(\gamma)} > 1~\TeV$  i.e.   $\langle E \rangle \simgeq 10~ \TeV$) nuclei energy losses
and re-acceleration are expected to be negligible (see e.g. \cite{Aharonian:00,Strong:04}).

Under the assumption of cylindrical symmetry the diffusion equation takes the simpler form \cite{Ptuskin:93}
 \begin{equation}\label{eq:diff2}
\left\{ -\frac{1}{r} \partial_r \left[ r D_\perp \partial_r \right] - \partial_z \left[ D_\perp \partial_z \right] + u_r \partial_r + u_z \partial_z \right\} N (E,r,z)  = Q (E,r,z)\;,
\end{equation}
where $D_{\perp}$ and $D_A$ are respectively the diffusion coefficient in the direction perpendicular to ${\bf B}_{\rm reg}$ and the antisymmetric (Hall) coefficient, while $u_r$ and $u_z$ are the drift velocities as defined in \ref{sec:appendix_num}. \Eref{eq:diff2} will be derived in \ref{sec:appendix_num}.

Several simulations of the diffusion coefficients in high turbulence MHD media have been  performed
\cite{Jokipii:99,Casse:02}. We adopt here the expressions for these coefficients which have been derived in \cite{Candia:04} and used in \cite{Candia:05}.
Those expressions provide $D_\bot$ and $D_A$ as functions of the regular magnetic field and of the {\it turbulence level} $\sigma$ in that point. Respect to other works, where the CR density has been simulated by using a mean value of the diffusion coefficients as derived from the observed secondary/primary ratio of CR nuclear species, our approach offers the advantage to provide the diffusion coefficients {\it point-by-point}. This allows to test how the expected neutrino and $\gamma$-ray emissions depend  on the properties of the turbulent components of the GMF. We verified, 
by means of dedicated runs, that the simulated escape time of nuclei found with our numerical code is compatible with that estimated from observations of the B/C ratio measured at energies around the GeV (see e.g. \cite{Strong:04}).

All the advantages of getting a realistic spatial distribution of CR by solving the diffusion equation, however, were not fully exploited in \cite{Candia:05}. In that work, in fact, the author solved equation (\ref{eq:diff2}) analytically, which was possible only considering an over-simplified spatial dependence of the diffusion coefficients. For example, $D_\bot$ was assumed to be spatially homogeneous.\\
In this work we solve numerically\footnote{The code is based on a previous Fortran version wrote by J. Candia which we translated in {\tt C}\begin{scriptsize}++\end{scriptsize} and improved in several parts.} (\ref{eq:diff2}) under more general conditions. We impose the boundary conditions $N = 0$ on a the cylindrical surface $(r = 30~\kpc, z = z_t$) where the turbulent halo is supposed to end and CRs escape to infinity without further diffusing.
For any given value of the particle rigidity the code provides a bi-dimensional histogram $\psi(r, z)$ mapping the particle spatial distribution in the stationary limit.  

\subsection{Models}
In order to verify how much the large uncertainties in the knowledge of the GMF properties may affect
our predictions for the CR distribution, we consider and test several models. 

Our first result of these tests is that the CR spatial distribution and spectrum are almost independent of the radial and vertical lenght-scales of the regular magnetic field.  Therefore the large observational uncertainties in $r_B$ and $z_r$ do not affect our final results. In the following we will always assume $r_B = 8.5~\kpc$ and $z_r = 1.5~\kpc$. 

More significant is the effect of changing the turbulent halo scale height $z_t$ as that parameter determines the length over which CR have to diffuse before escaping to infinity (here we always assume that $z_r < z_t$, as suggested form observations; see \sref{subsec:mf}). Indeed  we found that in the inner Galaxy, i.e. where sources are most abundant, the CR density increases with $z_t$.  We considered several values of that parameter finding that  $z_t = 3~\kpc$ 
provide the best matching of EGRET data (see \sref{sec:experiments}). 

 In \tref{tab:models} we summarise the main features of all models that will be discussed in this paper.
\begin{table}[!ht]
\caption{\label{tab:models}The main properties of the models considered in this section.}
\begin{indented}
\item[]\begin{tabular}{@{}ccccc}
\br
model \# & SNR & $\sigma(r)$ & turbulence & $z$-symmetry \\
\mr 
0 &  CB \cite{Case:1998qg} & 1  & Kolmogorov & S \\
1 & Ferriere \cite{Ferriere} & 1  &  Kolmogorov & S \\
2 & Ferriere \cite{Ferriere}  & like SNR &  Kolmogorov  & S \\
3 & Ferriere \cite{Ferriere}  & 1  & Kraichnan & S \\
4 & Ferriere \cite{Ferriere} & 1 &  Kolmogorov & A \\
\br
\end{tabular}
\end{indented}
\end{table}

\begin{figure}[!th]
 \centering
 \includegraphics[scale=0.78]{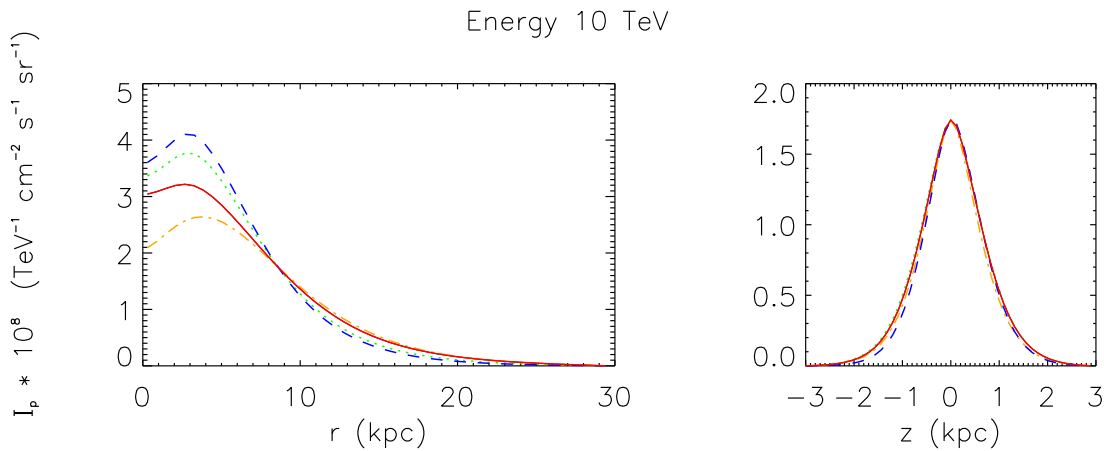}
 \caption{The proton flux profiles at $E = 10~\TeV$ are plotted along the galactic midplane
(left panel) and along the $z$-axis  at $r = r_\odot$ (right panel).  The dash-dotted line
 (orange),  continuos (red), dotted (green), and dashed (blue) correspond to the  models 0,2,1,3 respectively (see text). All fluxes have been normalised to  the observed value at the Sun position.}
 \label{fig:CRprof_le}
\end{figure}

\begin{figure}[!th]
 \centering
 \includegraphics[scale=0.78]{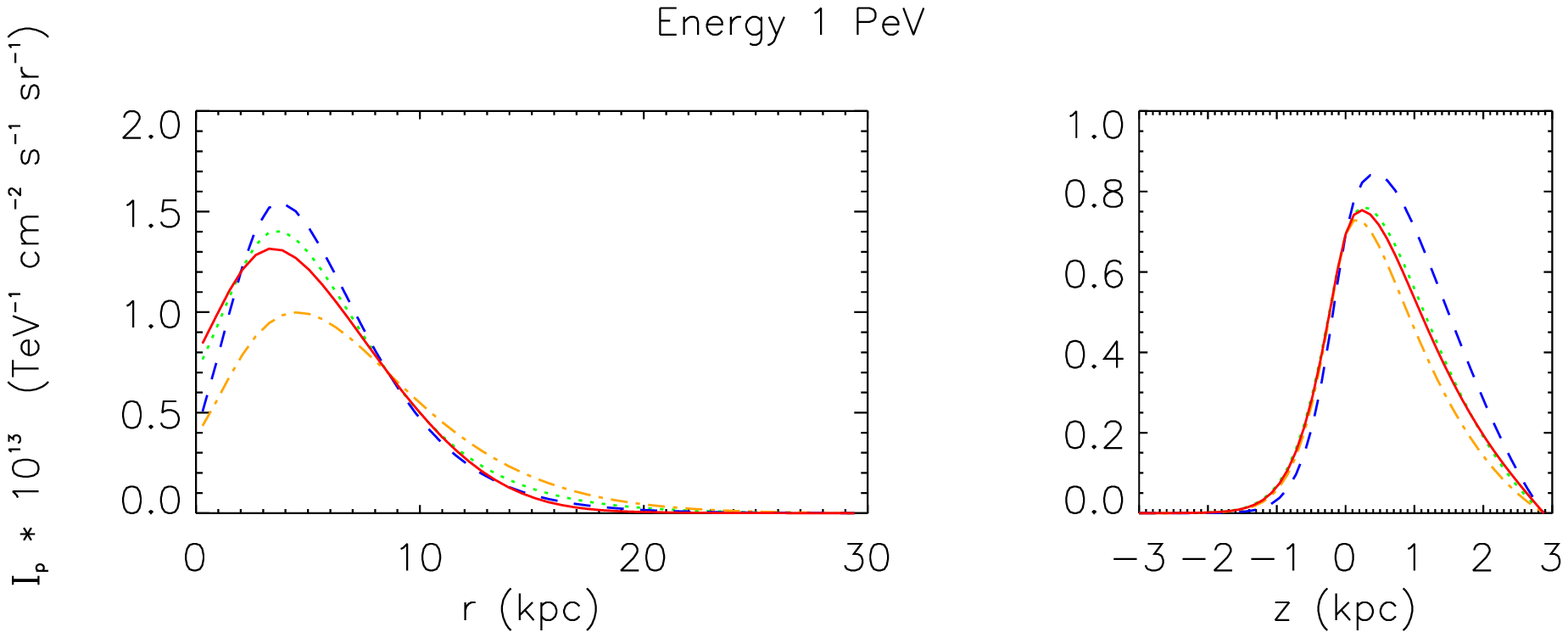}
 \caption{The same as \fref{fig:CRprof_le}  for $E = 1~\PeV$.}
 \label{fig:CRprof_he}
\end{figure}

In \fref{fig:CRprof_le} and \ref{fig:CRprof_he} we show the high energy proton fluxes along two significant sections of the galactic halo: the galactic plane and a plane perpendicular to it at $r = r_\odot$.
The effect of changing the SNR distribution is evident when comparing models 0 and 1 as both have the same GMF turbulent spectrum (Kolmogorov) and strength ($\sigma = 1$ everywhere on the GP). It is evident that the adoption of the SNR distribution given in \cite{Ferriere} gives rise to  a $\sim 30~\%$ increase in the CR density in the GC region with respect to that derived following \cite{Case:1998qg}. 

We also investigate the effect of changing the turbulence level uniformly in the magnetic halo finding a marginal effect at energies below the PeV.
More interesting is the effect of assuming a radially dependent turbulence strength $\sigma(r)$.
In model 2 we assume that quantity to follow the same radial profile of SNR as shown in the upper curve of \fref{fig:SNR}. That choice is justified by the well known argument according to which MHD turbulence in the ISM  is powered by SN ejecta.
By comparing proton flux profiles in \fref{fig:CRprof_le}, \ref{fig:CRprof_he} for the models 1 and 2 (as both are obtained by using the same value of $\sigma = 1$ at the solar circle) we find that a radially dependent $\sigma$ gives rise to a smoothing of the CR density distribution with respect to that of sources. That is to be expected as regions which are poor of sources are more easily filled by CR coming from more active regions if the turbulence strength is locally smaller. The effect is less evident at higher energies since in that
case CR escape more rapidly along the $z$ axis.

In the above we assumed a Kolmogorov power spectrum (i.e. $\gamma = 5/3$) for the turbulent component of the GMF. It is interesting to investigate how the CR spatial distribution is affected by adopting a different spectrum. A reasonable possibility is to assume a Kraichnan spectrum (model 3) which is characterised by $\gamma = 3/2$.  Noticeably, in this case $D_\perp \propto  E^{1/2}$ so that the value of the spectral index at the sources, which is required to fit the observed proton spectrum at the Earth, is $\beta = 2.7 - 0.5  = 2.2$.  This is in good agreement with the value measured for the $\gamma$-ray spectrum of several SNRs (see e.g. \cite{Beacom:06} and ref.s therein).
 Since the Kraichnan spectrum is harder than Kolmogorov's, it gives rise to a more tight confinement of CR in the nearby of sources explaining the higher density in the inner Galaxy which we observe for that model in \fref{fig:CRprof_le}, \ref{fig:CRprof_he}.

The deformation of the vertical profiles observed for all models at high energy is due to the
Hall diffusion which gives rise to a drift in a direction perpendicular to both ${\bf B}_{\rm reg}$
(along ${\hat \phi}$) and the radial component of ${\bf \nabla \Phi}_p$.

Finally, we also considered the effect of changing the symmetry of the regular magnetic field with respect to the galactic plane (model 4). In all previous cases a symmetric (S) configuration (i.e. ${\bf B}_{\rm reg} $
does not reverse its sign for $z \rightarrow - z$) was considered. Since observations did not settle what the actual symmetry is yet, it is interesting to consider the possible implications of its change. In \cite{Ptuskin:93,Candia:02} it was already shown that if the magnetic halo is A-symmetric, Hall diffusion gives rise to a significant increase of the CR density in the GC region at energies above the PeV.   
We confirm the existence of that effect. The consequences for the neutrino spectrum where investigated in \cite{Candia:05}. However, as they are negligible in the energy range considered in this work we will disregard  the A-symmetric models in the following.

As we mentioned, the diagrams in this sections refer to protons only. The corresponding flux profiles for composite nuclei can be obtained by a simple energy rescaling  $(E \rightarrow E/Z)$ and by using the proper normalisation  given by the observed flux of each nuclide.
Therefore,
 \begin{equation}
\frac{dn_{Z}(E;~r,z)}{dE} = \sum_{Z}  I_p(E_0) ~f_{Z} \left( \frac{E}{E_0} \right)^{-\alpha_Z}~ \psi (E/Z; r,z)
\label{nZ}
\end{equation}
where $f_{Z}$ and $\alpha_Z$ are respectively the locally observed fraction at 1 TeV of the species with charge $Z$  with  respect to protons  and their spectral slope. Recent compilations of measurements of both quantities can be found in \cite{Horandel:03}.
We conclude this section by noticing that although the CR energy spectrum is not exactly the same over the whole galaxy, this has negligible implications for the following results. 
 
\section{Mapping the $\gamma$-ray and neutrino emission}\label{sec:neutrinos}

We start by considering $\gamma$-ray production by the decay of neutral pions which are generated by the interaction
 of the nucleonic component of CR with the ISM (mainly hydrogen and helium).
Here we do not consider a possible contribution to the diffuse $\gamma$-ray flux which may be originated
by Inverse-Compton (IC) scattering of relativistic electrons with the background radiation (see e.g. \cite{Aharonian:00,Strong:98}).  Bremsstrahlung  is negligible at the energies considered in this work.
Due to the low density of the ISM practically all mesons decay before interacting with matter.
Since at those high energies constituent nucleons interact independently one from the other, here we need only to consider elementary inelastic nucleon-nucleon scattering. Furthermore, we can safely assume proton-neutron invariance.
In the energy range considered in this work $(E_\gamma < 100~\TeV)$, the attenuation of the 
$\gamma$-ray flux due to pair-production scattering onto the CMB radiation can be  neglected
(see e.g. \cite{Berezinsky:93,Ingelman:96}).

High energy neutrinos $(E > 1~\TeV)$  are prevalently originated by the decay of charged pions and kaons which are generated by the same hadronic scattering process considered for the production of $\gamma$-rays.
At the source only electron and muon neutrinos are produced.
During the propagation over galactic distances neutrino oscillations redistribute equally the  neutrino budget among all 
lepton families. In the following we will be interested only in the $\nu_{\mu} + {\bar \nu}_\mu$ flux reaching the 
Earth.

\begin{figure}[!th]
 \centering
 \includegraphics[scale=0.5]{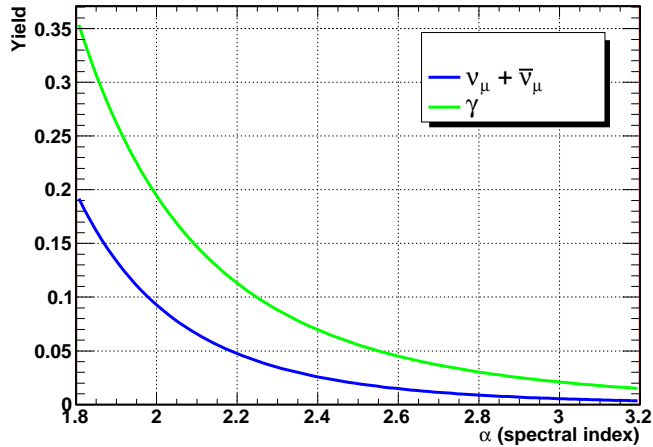}
 \caption{The $\gamma$-ray  and ${\nu_{\mu} + {\bar \nu}_\mu} $ yields are 
 represented as a function of the primary nuclei spectral index $\alpha$. Neutrino oscillations are accounted 
 for.} 
 \label{fig:yields}
\end{figure}

Under the assumption that the primary proton spectrum is a power-law and that the differential cross-section follows a scaling behaviour (which is well justified at the energies considered in this  paper), the $\gamma$-ray (muon neutrino) emissivity, can be written as (see e.g. \cite{Berezinsky:book,Gaisser:book})
\begin{equation}
 Q_{\gamma ~(\nu_{\mu} + {\bar \nu}_\mu)} (E_\nu; r,z) =  f_N 
 \frac{dn_p(E_\nu, r,z)}{dE}~ \sigma_{pp} c \; n_H(r,z) \; Y_{\gamma~(\nu_{\mu} + {\bar \nu}_\mu)} (\alpha)\;.
 \label{qnu} 
 \end{equation}
 Here $\alpha$ is the primary CR spectral index and  $\sigma_{pp} \simeq 3.3  \times 10^{- 26}~\cm^{-2}$ is the 
 $pp$ inelastic cross-section at $1~\TeV$ \cite{pdg:06}.
The $\gamma$-ray  (muon neutrino) yield $Y_ {\gamma~(\nu_{\mu} + {\bar \nu}_\mu)} (\alpha)$, as determined in \cite{us:05}, is shown in \fref{fig:yields}\footnote{An almost  $10\%$ contribution to the photon emissivity coming from $\eta$ decay, which was not considered in \cite{us:05}, has been included here.}. Those values are  in agreement with previous results (see e.g. \cite{Vissani:04,Lipari:06,Kelner:06}) within 20\%.
Above few GeV's, yields are practically independent of energy.
For $\alpha = 2.7$ the $\gamma/\nu_{\mu} + {\bar \nu}_\mu$ flux ratio is $3.1$.

The factor $f_N$ in (\ref{qnu}) represents the contribution from all other nuclear species both in the CR
and the ISM. We find
\begin{equation}
f_N \simeq 
\sum_{Z,A} f_{Z} \frac{\alpha_1}{\alpha_Z}\left(\frac{E_\nu}{E_0}\right)^{\alpha_1-\alpha_Z} 
\left(1+4\frac{n_{\rm He}}{n_{\rm H}}\right)A^{1 -\alpha_Z} \simeq 1.4 ~,
\end{equation} 
which is in good agreement with the value found in other works (see e.g. \cite{Dermer:86,Mori:97}).
Here we used experimental values of $f_{Z}$ and $\alpha_Z$ as given in \cite{Horandel:03} and the helium/hydrogen ratio in the ISM ${n_{\rm He}}/{n_H}  \simeq 0.09$ \cite{Ferriere}. All other nuclear components in the ISM give a negligible contribution.

The differential $\gamma$-ray (neutrino) flux reaching the Earth is given by the line integral
\begin{equation}
\label{line_int}
I_{\gamma (\nu_{\mu} + {\bar \nu}_\mu)} (E_\nu;~b,l) = \frac{1}{4\pi} \int
Q_{\gamma~(\nu_{\mu} + {\bar \nu}_\mu)} (E_\nu;~b,l,s)~ {\rm d}s ~,
\end{equation}
where $s$ is the distance from the Earth and $(l,b,s)$ are related to $(r,z,\phi)$ through
\begin{equation}
z = s \sin b \\
r = \sqrt{(s \cos b \cos l - r_\odot)^2 + (s \cos b \sin l)^2}~,
\end{equation}
in cylindrical symmetry.
Finally, the integrated fluxes are determined by integrating the power law spectrum over the energy up to 1 PeV.

For the sake of clarity, in the following we will show flux diagrams as obtained only with our model 3B (model 3 for the CR distribution and B for the gas) which is our preferred model. As we mentioned in \sref{subsec:gas}, the gas model B gives the best matching of CO surveys, while the CR model 3 has to be preferred because, by adopting gas model B, it best reproduces EGRET observations above few GeV's (see \sref{subsec:gammaexp}).
At the end of this section we will briefly discuss how our predictions would change adopting different models.

\begin{figure}[!th]
 \centering
 \includegraphics[scale=0.78]{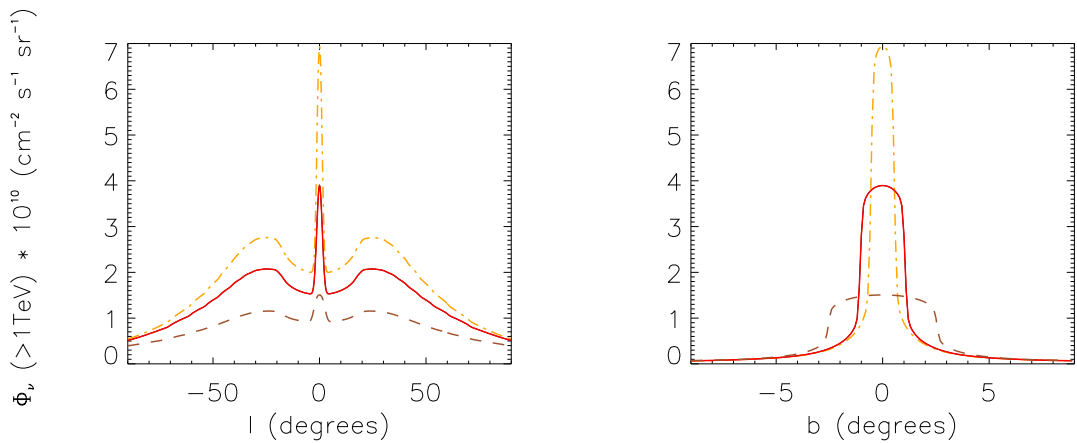}
 \caption{The integrated $\nu_\mu + {\bar \nu}_\mu$ flux is shown along the galactic equator ($b = 0$, left panel) and along $l = 0$ (right panel). Fluxes are averaged over $1^\circ \times 1^\circ$ (dot-dashed, orange line), $2^\circ \times 2^\circ$ (continue, red line), $5^\circ \times 5^\circ$ (dashed, brown line) angular bins. The corresponding $\gamma$-ray flux can be obtained by multiplying this diagram by 3.1.}
\label{fig:nuprof}
\end{figure}

In \fref{fig:nuprof} we show two representative sections of the  neutrino flux profile above 1 TeV . 
In order to show how the expected signal may depend on the experimental angular resolution, in the same figure  we draw  the  flux averaged  over angular bins of different sizes. It is evident that due to the narrowly peaked behaviour of the gas density along the GP ($b = 0$), the averaged flux which may be measured from these regions should change significantly by varying the angular resolution. 

\begin{figure}[!ht]
 \centering
 \includegraphics[scale=0.78]{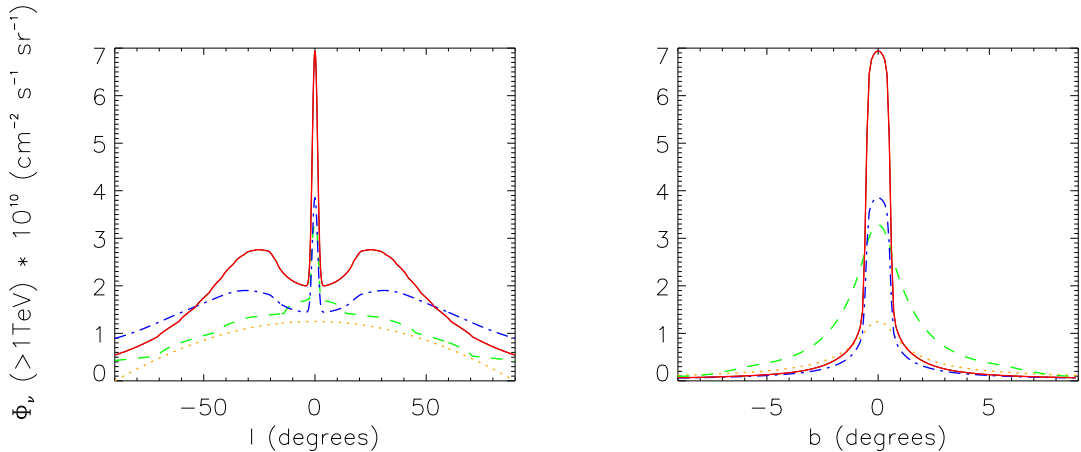}
 \caption{The $\nu_\mu + {\bar \nu}_\mu$ flux for $E > 1~\TeV$, averaged over $1^\circ \times 1^\circ$ angular bins, is shown. The continuos (red), dot-dashed (blue), dashed (green), and dotted (orange) curves correspond respectively to: our model 3B, a model with the same gas density distribution (model B) but a uniform
 CR density as observed at the Earth position;  the model considered in  \cite{Berezinsky:93};  
 the model considered in \cite{Ingelman:96}. The corresponding $\gamma$-ray flux can be obtained by multiplying this diagram  by 3.1.}
 \label{fig:nuprof_comp}
\end{figure}

In \fref{fig:nuprof_comp} we compare our results with those obtained in \cite{Berezinsky:93} and \cite{Ingelman:96} which have been derived assuming a uniform CR density. We also show the flux as obtained by using the same gas model B  but a uniform CR density. By comparing those profiles the reader can see as the flux that we expect from the central region of the GP is significantly larger than in \cite{Berezinsky:93}. That  discrepancy is due, for a large fraction, to the SNR  overdensity in the molecular ring region\footnote{The reader should also note that in \cite{Berezinsky:93} the  $N_H  \rightarrow \Phi_{\gamma} (E > 1~\TeV;~b,l)$ conversion factor is $6.02\times 10^{-33}$ while we have $8.87 \times 10^{-33}~\left(\TeV~\s~\sr \right)^{-1}$.}.

\begin{figure}[!ht]
\centering
\includegraphics[angle=90, scale=0.6]{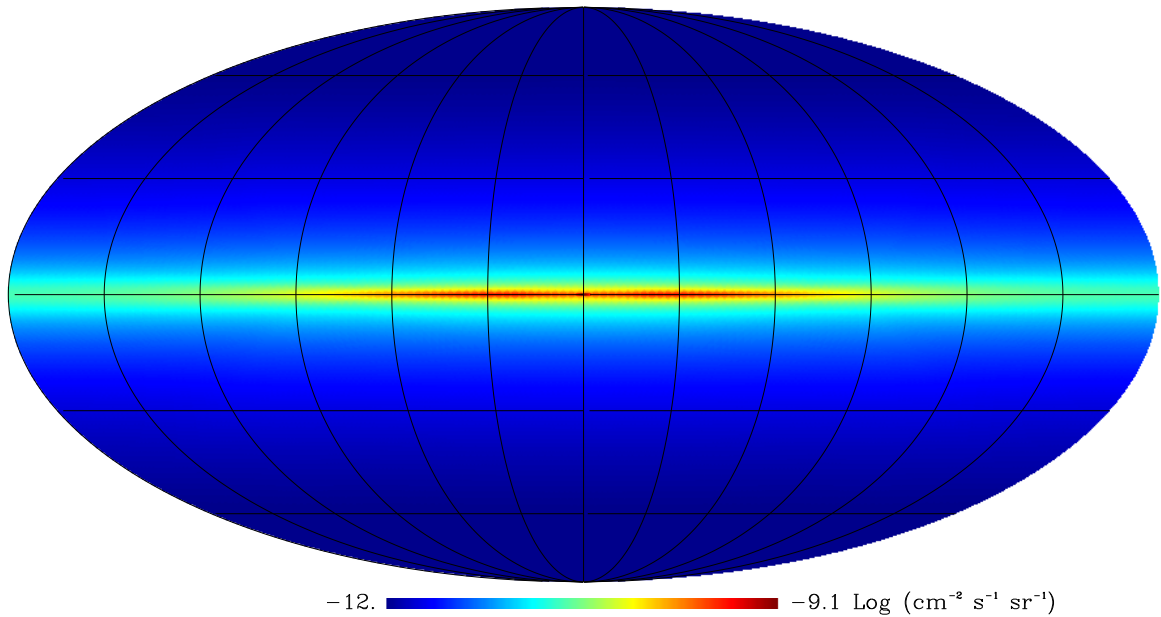}
\caption{The all-sky map of  $\Phi_{\nu_{\mu} + {\bar \nu}_\mu} (E_\nu > 1~\TeV)$ is shown in  galactic coordinates (the GC is in the center). This map has been obtained with HEALPix \cite{Healpix}. The flux distribution is  projected in such a way that all pixels in this map correspond to the same solid angle. The corresponding $\gamma$-ray flux can be obtained by multiplying this diagram by 3.1.}
\label{fig:numap}
\end{figure}

Finally in \fref{fig:numap} we represent a full sky map of $\Phi_{\nu_{\mu} + {\bar \nu}_\mu} (E_\nu > 1~\TeV;~b,l)$ obtained using model 3B. This figure has been done with the HEALPix package \cite{Healpix}\footnote{ See http://healpix.jpl.nasa.gov}.

We conclude this section by discussing how our predictions would change if different models
for the gas and CR distributions are adopted. We find that by using the gas model A rather than B, the
$\gamma$-ray and neutrino fluxes along the GP grow at most by a factor 2. We find comparable displacements by adopting any other of the CR models considered in \sref{sec:diffusion}. Therefore, we conclude  that, under the hypothesis on which this analysis is based, our results are very little model dependent.

\section{Comparison with available observations}\label{sec:experiments} 

\subsection{Gamma-rays} \label{subsec:gammaexp}

First of all, we would like to compare our results  with EGRET observations.
That is required to put our results onto  more solid grounds and to verify that our choice of the preferred  model 
is consistent with those observations.  Here we consider only  EGRET measurements of the diffuse emission along  the 
GP between 4 and 10 GeV \cite{Cillis:05,egret}. Such a comparison requires a considerable  extrapolation of our 
previous  findings. We are allowed to do that  since, already for  $< 10~\TeV$,  nuclei propagation takes place deep into the spatial diffusion regime so that the energy dependence of the diffusion coefficients do not change going to lower energies (see e.g. \cite{Casse:02,Candia:02}).   As a consequence, in the stationary regime the CR distribution above  10 GeV  is given by a uniform  $\propto E^{-2.7}$  rescaling of that shown in \fref{sec:diffusion}.
We neglect small corrections due to the variations of the $pp$  scattering cross section with the energy.  
 
\begin{figure}[!ht]
\centering
\includegraphics[scale=1.0]{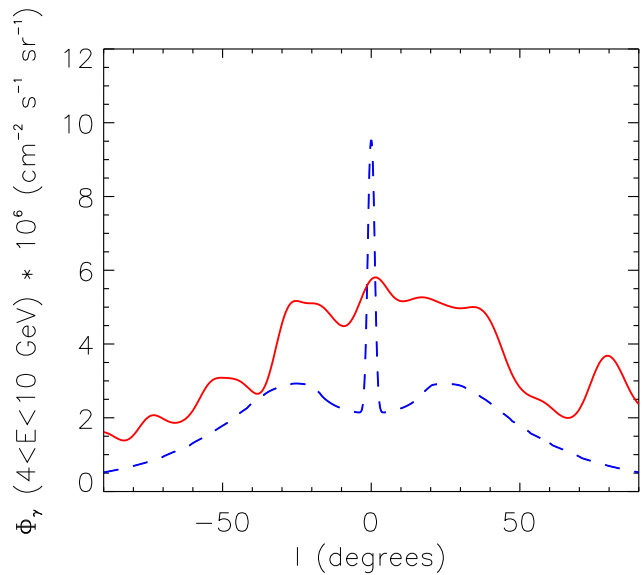}
\caption{The profile of  $\gamma$-ray flux  along the Galactic Plane, between 4 and 10 TeV, as obtained with our reference  model 3B ( blue, dashed line) is compared with that measured by EGRET (red, continuos line). Fluxes are averaged over $0.5^\circ \times 0.5^\circ$ angular bins.}
\label{fig:egret_match}
\end{figure}

\begin{figure}[!ht]
\centering
\includegraphics[scale=0.7]{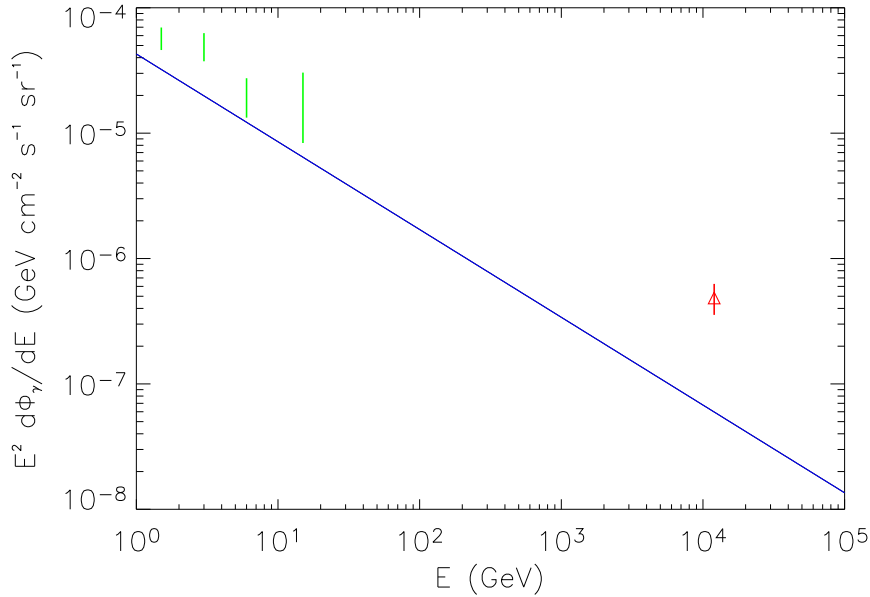}
\caption{The blue (continuos line) is the differential $\gamma$-ray spectrum multiplied by $E^2$  
 as obtained with our reference  model 3B in the region $73.5 ^\circ \le  l \le 76.5^\circ,\ |b| \le 1.5^\circ$ . 
 Green (red)  bars represent EGRET (MILAGRO) measurements in the same window.}
\label{fig:spectrum}
\end{figure}

In  \tref{tab:comp} and in \fref{fig:egret_match} we compare respectively theoretical mean  fluxes and flux 
profiles along the GP   with EGRET measurements \cite{egret}.  
Considering the large  uncertainties involved in the modelling of the CR and gas distribution, the agreement is rather good.   It should be noticed that in the energy range $4-10~\GeV$  the  IC contribution to the
 photon flux from the GP is expected to be subdominat  (see e.g. \cite{Aharonian:00,Strong:04}). 
In \fref{fig:spectrum} we also show our prediction for the  $\gamma$-ray differential spectrum between 1 GeV and 100 
TeV  in the region $73.5 ^\circ \le  l \le 76.5^\circ,\ |b| \le 1.5^\circ$ and we compared it
with EGRET \cite{egret}  and MILAGRO measurments \cite{Abdo:2006} in the same region 
\footnote{The best fit of the spectral slope observed by EGRET above 1 GeV is $-2.6$ \cite{Hunter:97}. 
 That  is compatible with that adopted in our paper within errors.} .  
All the theoretical fluxes shown in this section have been derived using the CR model 3 and gas model B. 
As we mentioned in the previous section, while we prefer gas model B because it provides the best description of CO surveys, once made that choice,  models 3 is that which gives the best fit of  EGRET data.  Anyhow, passing from one model to another,  our predictions change  by a factor 2 at most. 

Reassured by these findings we are now ready to compare our results with measurements performed at higher 
energies.  Several air shower array  experiments looked for the $\gamma$-ray diffuse emission of the Galaxy 
above the TeV.  The most interesting results are those of TIBET \cite{Tibet:obs} and MILAGRO \cite{Milagro:obs,Abdo:2006}. Both experiments probed different regions of the GP.
In all those regions only upper limits were found but in Cygnus  where MILAGRO 
found a significant excess on the background \cite{Milagro:obs,Abdo:2006}.
In \tref{tab:exp_comp} we compare our predictions, as obtained with  our preferred 3B model
(see \sref{sec:neutrinos}), with those measurements. 
With the exception of Cygnus (see also the discussion about the GC ridge at the the of this subsection)
in all other regions  we predict  fluxes which are significantly below the experimental limits.
Therefore, there is still  room  for a large  IC contribution.

\begin{table}[!ht]
\caption{\label{tab:exp_comp} In this table our predictions for the mean $\gamma$-ray flux in some selected regions of the sky are compared with some available measurements. Since measurements' errors are much smaller than theoretical uncertainties they are not reported here.}
\label{tab:comp}
\begin{indented}
\item[ ]
\begin{tabular}{@{}cccc}
\br
sky window  & $E_{\gamma}$ & \multicolumn{2}{c}{$\Phi_\gamma(>E_\gamma)~(\cm^2~\s~ \sr)^{-1}$}\\
\mr
            &             &   our model & measurements \\
\mr
$|l|  < 10^\circ,\ |b| \le 2^\circ$ & $4~\GeV$ & $ \simeq  4.7 \times 10^{-6}$ &$\simeq  6.5\times 10^{-6}$ \cite{egret}\\
\mr 
$20^\circ \le  l \le 55^\circ,\ |b| \le 2^\circ$ & $3~\TeV$ & $\simeq 5.7 \times 10^{-11}$ & $\le 3 \times 10^{-10}$  \cite{Tibet:obs} \\
 & $4~\GeV$ & $\simeq  4.4 \times 10^{-6}$ & $\simeq  5.3 \times 10^{-6}$  \cite{egret}\\
\mr
$73.5 ^\circ \le  l \le 76.5^\circ,\ |b| \le 1.5^\circ$ &  $12~\TeV$ & $\simeq   2.9\times 10^{- 12}$ & $\simeq  6.0\times 10^{-11}$  \cite{Abdo:2006}\\
 &$4~\GeV$ & $\simeq  2.4 \times 10^{-6}$ & $\simeq  3.96 \times 10^{-6}$ \cite{egret} \\
\mr
$140^\circ < l < 200^\circ, \ |b| < 5^\circ$ & $3.5~\TeV$ & $\simeq  5.9 \times 10^{-12}$ & $\le 4 \times 10^{-11}$  \cite{Milagro:obs}\\
 & $4~\GeV$ & $\simeq  5.9 \times 10^{-7}$ & $\simeq 1.2\times 10^{-6} $  \cite{egret} \\
\br
\end{tabular}
\end{indented}
\end{table}

Concerning the excess in the Cygnus region observed by MILAGRO, we confirm the 
conclusion \cite{Abdo:2006,Fields:06}  that it cannot be explained by the interaction of the diffuse component of galactic CR with the gas in that region  (see \fref{fig:spectrum}).
Indeed, we find that a CR local over-density of  about 20 is required to explain that signal in terms of hadronic emission.

A similar effect has to be invoked to explain HESS measurement of the diffuse $\gamma$-ray TeV emission from the GC ridge region \cite{Hess2006}. In fact, even accounting for the detailed 3-dimensional gas model recently developed in \cite{Ferriere:07} for the GC region, we find that in the region $|b| < 0.8^\circ, |l| < 0.3^\circ$ the expected  $\gamma$-ray flux for $E > 1~\TeV$ is  $\sim 2.4 \times 10^{-9}~\cm^{-2}~\s^{-1}~\sr^{-1}$, which is almost 10 times smaller than HESS measurement. It is worth to notice that the slope of the spectrum measured by HESS is close to $-2.3$ which is quite different from that of galactic CR. That probably means that primary particles in the GCR have a local origin. It is evident that the analysis performed in this work, as well as in related papers \cite{Strong:98,Aharonian:00,Strong:04}, cannot account for those kind of local emissions.

\subsection{Neutrinos}\label{subsec:nu_exp}

The only available upper limit on the neutrino flux  from the Galaxy has been obtained 
by the AMANDA-II experiment  \cite{Kelley:05}.  Being located at the South Pole,
AMANDA cannot  probe the  emission from the GC.  In  the region  
$33^\circ < l < 213^\circ, \ |b| < 2^\circ$, and assuming a spectral index $\alpha = 2.7$, 
 their present constraint is 
\begin{equation}
I_{\nu_{\mu} + {\bar \nu}_\mu} < 6.6 \times 10^{-4} 
\left( \frac{E}{1~\GeV}\right)^{-2.7}~(\GeV~\cm^2~\s~\sr)^{-1}~,
\label{eq:limitAMANDA}
\end{equation}
which implies
$\Phi_{\nu_{\mu} + {\bar \nu}_\mu} (> 1~\TeV) < 3.1 \times 10^{-9} 
~(\cm^2~ \s~ \sr)^{-1}$. According to  our model the expected flux  in the same region is   
$\Phi_{\nu_{\mu} + {\bar \nu}_\mu} (> 1~\TeV) \simeq  4.2 \times 10^{-11} 
~(\cm^2~ \s~ \sr)^{-1}$. 

\section{The expected neutrino signal in the North Hemisphere}\label{sec:km3}

In the energy range $1-100~\TeV$ water (or ice) Cherenkov neutrino telescopes are best suited to look for up-going muons produced by charged-current interactions of muon neutrinos in the Earth. While the Earth offers almost a complete shielding from up-going atmospheric muons below the horizon, an unavoidable background is given by atmospheric neutrinos. Several estimates have been made of the atmospheric neutrino flux, based on different assumptions modelling  hadronic interactions (see e.g. \cite{Volkova,Agrawal:1995gk,Honda}).  Above the TeV all calculations almost agree predicting an averaged flux
\begin{equation}
F^{\rm atm}_{\nu_{\mu} + {\bar \nu}_\mu}(E_\nu) \simeq 4.6 \times 10^{-8}
\left(\frac{E_\nu}{1~\TeV}\right)^{-3.7} \TeV^{-1}~\cm^{-2}~\s^{-1}~\sr^{-1}
\label{eq:atmnuflux}
\end{equation}
though a $\sim 40~\%$ uncertainty remains due to the experimental error on the primary
CR spectrum and  the theoretical error modelling strange particle production.

Since, as it is evident from our previous results, the expected neutrino flux from the Galaxy is significantly smaller than such background, a suitable procedure has to be adopted to disentangle the signal. One possible approach is to search only for  neutrinos with $E \simgeq 100~\TeV$ \cite{Candia:05}. 
That however may be hard to do due to the very low flux at that high energies and to the Earth opacity
(only Earth skimming neutrinos or shower-like events produced by down-going neutrinos  can be detected in that case).
In  our opinion a  more promising  strategy is to search for up-going muon neutrinos above $1 - 10$ TeV. Their  arrival direction can be reliably reconstructed with an angular resolution as good as $0.5^\circ$ in water and that information may be used to identify the galactic emission as a localised excess of events.

While it is quite evident that the most natural  search  window  ({\it on-source region})  is  a narrow  band along the galactic equator (see \fref{fig:numap}),  its  optimal sizes 
have to be chosen by taking into account the angular extension of the source.  
In the previous section we showed that most of the neutrino flux from the Galaxy should be
concentrated  in the region $|l| < 50^\circ,\ |b| < 1^\circ$. 
Since the Gaussian width of the signal $ \sigma_{sig}\simeq 2^\circ$  is comparable to the expected 
angular resolution of neutrino telescopes, some attention has to be paid when choosing the 
latitude width of the search window. 
By assuming that both the line spread function of the experiment and the signal profile 
 along $b$ are Gaussian having widths $\sigma_{lsf}$ and $\sigma_{sig}$, respectively, 
the optimal  search window  width  is approximatively given by  (see e.g. \cite{Kelley:05}) $\displaystyle \Delta b  \simeq (1 - 2) \left( \sigma^2_{lsf} + \sigma^2_{sig} \right)^{1/2}$.
Since  for  a km$^3$  water NT the expected {\it line spread function} is $\sigma_{lsf}\simeq 
0.5^\circ$,  we think that  $\Delta b = 3^\circ$ is a reasonable value to adopt. 

Although to perform a detailed calculation of  the expected signal in a given NT is beyond the purposes of this work, here we perform a simplified estimate which is only intended to give the reader a feeling of the chances that forthcoming experiments have to achieve a positive  detection. 
 The expected muon detection rate  of neutrinos coming from that window is
\begin{equation}
{\dot N_\nu}(> E_\nu)  =   \int_{\Delta l} \int_{\Delta b} \int_{E_\nu}  dE \ A^{\rm eff}(E) ~v(b,l) ~\frac{dN_\nu (E;~l,b)}{dE_\nu}
\end{equation} 
where $A^{\rm eff}(E_\nu)$ is the effective area of the neutrino telescope and $v(l,b)$ is the visibility function (i.e. the fraction of time that a point in a sky with galactic coordinates $(l,b)$ spends above the visibility horizon). For example, $v(0,0) = 0.67$ at ANTARES geographical position.

Since the effective area is relatively weakly dependent on the arrival direction of the neutrinos, we adopt here a mean value obtained by averaging over all possible nadir angles. 
As a reference we use  $A^{\rm eff}(E_\nu)$ as  provided by the ANTARES collaboration 
\footnote{ Here we use an effective area as provided by the ANTARES collaboration for
point-like sources.  According to a recent analysis  \cite{Zornoza},   which 
adopts a suitable rejection strategy for the residual atmospheric muon background, 
the effective area for an all-sky diffuse flux differs very little from that for point-like 
sources. A dedicated analysis for the kind of emission considered in this work has not been performed yet.} (see e.g.\cite{Teresa}) and assume that for a  $\km^3$  experiment  it will  be 40 times larger. This is a rapidly growing function of the energy for $E \simleq 100~\TeV$.  
It is interesting to observe that  the product  $\displaystyle A^{\rm eff}(E) 
~\frac{dN_\nu (E;~l,b)}{dE}$, which is related to the detection efficiency, is peaked at 
about 1 TeV for  a $\alpha = 2.7$ power law spectrum.

\begin{figure}[!ht]
 \centering
 \includegraphics[scale=0.78]{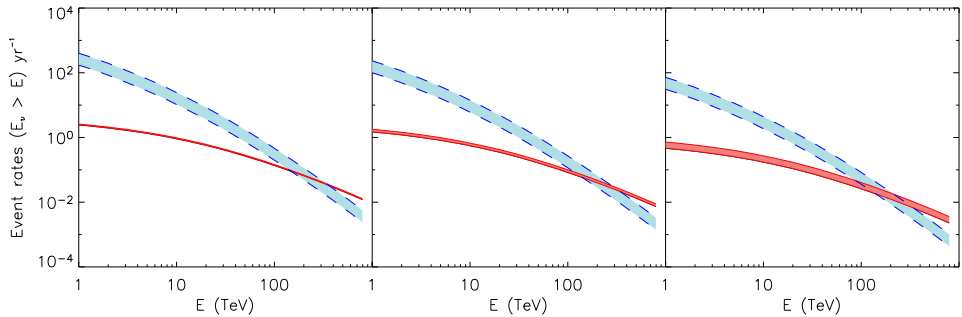}
\caption{The expected  ${\nu_{\mu} + {\bar \nu}_\mu}$ detection rate (red band, between continuos line)  
from the regions  $| l | <  50^\circ,\ |b| < 1.5^\circ$ (left), 
$| l | <  30^\circ,\ |b| < 1.5^\circ$ (centre), $| l | <  10^\circ,\ |b| < 1.5^\circ$ (right), 
is compared with the background due to atmospheric  neutrinos from the same region 
(ligth blue band, between dashed lines).  An effective area 40 times larger than ANTARES  has been adopted.
\ E is the neutrino energy. }
  \label{fig:nurates}
\end{figure}

In \fref{fig:nurates} we show the  integrated muon neutrino detection rate as a function of the minimum 
 energy to be expected in a $\km^3$ NT placed at the same geographical position of 
ANTARES.  The upper curve delimiting the signal band corresponds to model 3A, while the lower to model 2B. 
We compare that rate with the atmospheric neutrino detection rate expected in  ANTARES  \cite{Zornoza} 
multiplied by a factor 40 to account for  the larger effective area. 
Our calculation accounts for  the angular dependence  of the expected atmospheric neutrino  detection rate. 

According to \fref{fig:nurates}, the background is about 50 times larger than the signal above 1 TeV in the 
$|l| < 30^\circ$ region,  and about 5 times above 10 TeV. Therefore,
an excess may be detectable for  $E \simgeq 10~\TeV$ only after a considerable number of  years.

\section{Conclusions}\label{sec:conclusions}

We have performed a comprehensive calculation of the  neutrino and $\gamma$-ray flux distributions which are  expected to be originated by the interaction of CR nuclei with the interstellar gas.

Our computation  improves and updates  previous analyses  under several aspects.\\
Modelling  the spatial distribution of CR in the Galaxy, we accounted for a spatial dependence of the diffusion coefficients. That approach allowed us to test  how different  models of the turbulent component of the GMF
affect the CR distribution and the secondary $\gamma$-ray  flux  and to choose our preferred model  as that  which best reproduces EGRET $\gamma$-ray observations above few GeVs. Interestingly, we found that such a model gives also the best agreement with the hard $\gamma$-ray spectra observed  for several SNRs.\\
Having assumed that CR sources are distributed like SNRs, we adopted a detailed model of the spatial
distribution of those objects. \\Concerning the gas distribution, we considered  two  models, both for the HI and ${\rm H}_2$, which we  applied for the first time  in this framework. 
Although these models have significant differences in some regions, we showed that they give rise to neutrino and $\gamma$-ray fluxes which, when averaged over windows larger than few degrees squared, coincide  within a   factor of a 2.  We also found that  the  column density profile across the GP is almost the same for those two models, while it differs  significantly from that adopted in previous works \cite{Berezinsky:93, Candia:05}.  As a consequence, we predict the neutrino and $\gamma$-ray emissions to be more narrowly peaked along the GP. This improves significantly the perspectives to disentangle the signal from the almost isotropic background if  the experimental angular resolution will be better than $1^\circ-2^\circ$.

We compared our predictions  with the available observations. \\
Concerning  the $\gamma$-rays,  we found that above the TeV our predicted fluxes are below the experimental limits by less than an order of magnitude.  Since our analysis does not account  for a possible IC contribution, which may be not negligible above the TeV,  it is possible that those experiments will get a positive signal in the next few years.   The comparison of the experimental results with our predictions, as well as  with those of previous works (see e.g. \cite{Strong:98,Aharonian:00,Strong:04}), will help to understand which fraction of that emission is of  leptonic origin. \\Concerning the signal observed by MILAGRO in the Cygnus region \cite{Milagro:obs,Abdo:2006,Fields:06}, we confirm that it cannot be explained without invoking  a significant CR overdensity,  or a large IC contribution, which may be due to a local concentration of sources. 
A similar effect takes place in the GC region. We found that in order to explain the excess observed by HESS
from the GC ridge \cite{Hess2006} the CR flux must exceed  our prediction by a factor of about 10.  
This is not unexpected since the distribution of  star forming regions, hence of CR sources, is known to be
quite clumpy. It is understood that the kind of simulation performed in this work, as well as in previous ones \cite{Strong:98,Aharonian:00,Strong:04},  can only  model the mean CR distribution smoothed on scale of several hundred parsecs.  

We also extrapolated our results down to few GeVs, where the IC scattering emission is expected to be subdominant. We found that  in the most dense  regions the flux distribution agrees reasonably  well with EGRET observations.   This result encourages us to improve the accuracy of our analysis
by using, for example, more detailed models of the gas distribution,   so to be able  
 to simulate what GLAST may observe above the GeV  at least in some limited regions of the sky.   

Going back to neutrinos, we  compared our predictions for the muon neutrino flux from the GP
with the experimental limit recently established by  AMANDA-II \cite{Kelley:05}. 
We found  that our predicted flux is almost two orders of magnitude below that limit. Unfortunately, also  ICECUBE \cite{Icecube} will hardly be able  to get a positive signal.
\\Since a neutrino telescope placed in the North hemisphere may have better changes, we investigated this possibility in some details.   Assuming that such an instrument will be placed at the same position of ANTARES  \cite{Antares} and have a  40 times larger  effective area we estimated the expected signal and the background along the Galactic Equator. We found that the detection of the smooth component of the diffuse emission may require more than 10 years of data taking. It should be noted, however, that our analysis provides only a lower limit to the expected emission from the Galactic plane. As we mentioned in the above, several observations suggest that the CR and gas distributions may be more clumpy than what considered in this work. Furthermore, fluctuations of those quantities are likely to be spatially correlated. This may lead to a significant enhancement of the neutrino flux from some regions as may be the case for the GC ridge \cite{us:05,Beacom:06}  and  the Cygnus region \cite{Beacom:07,Ancho:07}. In a forthcoming paper we will try to develop our analysis in order to model at least some of these  effects.

\section*{Acknowledgements}

We thank J. Candia for valuable discussions and for providing us with the FORTRAN
code which was used in \cite{Candia:02}  to solve the cosmic ray diffusion equation.
We are grateful to H. Nakanishi for sending to us unpublished results of his analysis on the HI density distribution
and thank D. Gaggero for comparing our results with those obtained using a  3D gas distribution in the galactic bulge region. We thank V. Cavasinni and  V. Flaminio for their encouragement and for reading 
this manuscript before its submission. We also thank, R. Aloisio, C. Baccigalupi, R. Bandiera, P. Blasi, A. Celotti, S. Degl'Innocenti, K. Ferriere, F. Halzen,  J. Kelley, J. Kirk, T. Montaruli, for useful discussions and comments.
Finally we thank the referee for constructive criticisms.
D. G. and L.M. are members of the ANTARES collaborations. D.G. is also affiliated to the GLAST
collaboration. Most  of the work by C.E. was done at the Universit\`a di Pisa and INFN, Pisa in the ANTARES group. Some of the results presented in this paper have been obtained using the HEALPix \cite{Healpix} package.

\appendix
\section{Models of the gas distributions}
\label{sec:appendix_gas}

The HI and ${\rm H}_2$ density distributions are generally inferred from radio observations of the $21~\cm$ line and $^{12}$CO rotational line emissions respectively. The galacto-centric radial density profiles are estimated by converting the line-of-sight velocity (which is determined from the Doppler shift) into heliocentric distance by means of the Galactic rotation curve (see e.g. \cite{Ferriere}).
This operation is impossible for the whole Galaxy and it always involves large uncertainties so that some phenomenological models have to be invoked.

We are going here to describe in detail the models sketched in \sref{subsec:gas}.
\begin{description}
\item[Model A] It has been developed by Nakanishi and Sofue (NS) in \cite{Nakanishi:03} for the HI and by the same authors in \cite{Nakanishi:06} for the  ${\rm H}_2$. 
For both gas components NS developed  3-D models ($r,z,\phi$) which, however, do not  cover the entire Galaxy. By averaging  over the azimuth they get 2D-distributions which are approximatively symmetric with 
respect to the galactic plane. NS assumed that the vertical profiles of the HI and ${\rm H}_2$ densities have the 
form  $\displaystyle n(r,z) = n(r,0)\;  {\rm sech}^2 \left\{ \log (1 + \sqrt{2}) \;  z/z_{1/2}(r)\right\}$ which is
 almost coincident with a Gaussian peaked at the galactic plane half-width $h(r) = z_{1/2}(r)/\sqrt{\ln(2)}$. The
  quantity $2\cdot z_{1/2}(r)$  is usually called the Full Width to Half Maximum (FWHM) scale height. Then, 
  they derived binned radial distributions for $z_{1/2}(r)$ and the gas density $n(r,0)$ along the Galactic Plane.
\item[Model B] It is the product of the combination of results of different analyses which have been 
separately  performed  for the disk and the galactic bulge. For the ${\rm H}_2$ and HI distributions in the bulge we use a  detailed 3D model recently  developed by Ferriere et al. \cite{Ferriere:07} on the basis of several observations. In that model both the ${\rm H}_2$  and the HI are concentrated in two main structures. The  central molecular zone  (CMZ) appears as a quite dense $~500 \times  30~\pc$ wide (sizes at half maximum density) ellipse on the plane of the sky (for the HI the vertical extension is 90 pc) containing almost $2 \times 10^7~M_\odot$ in  ${\rm H}_2$ and $1 \times 10^6~M_\odot$ in HI. It gives rise to a pronounced peak in the gas column density in the direction of  the GC. The holed Galactic bulge disk (GBD) is  3 kpc long and 1 kpc wide  toroidal  structure. Its mass is comparable to that in the CMZ ($\sim 3 \times 10^7~M_\odot$ in  ${\rm H}_2$ and $3 \times 10^6~M_\odot$ in HI) but it is spread over a much larger volume  so that the gas density in that region is much smaller than in the CMZ. Furthermore, the GBD is significantly inclined with respect to the plane of the sky so that its contribution to the gas column density along the line of sight is quite small.

For the molecular hydrogen in the disk we use the Bronfman's et al. model \cite{Bronfman:88}. Although such a model is almost two decades old, it was shown \cite{Ferriere} that it still provides a good description of recent CO surveys \cite{Dame:01}. Once averaged over the azimuth, Bronfman's et al. distribution is well approximated by a disk with a Gaussian vertical profile symmetric with respect to the Galactic plane. Since in \cite{Bronfman:88} $r_\odot = 10~\kpc$ was adopted, we correct the gas densities and the scale heights given in that paper to make them compatible with the value  $r_\odot = 8.5~\kpc$ we use in this work.

Furthermore, we accounted for the different values of the  ${\rm H}_2$-CO conversion factor which is $X = 0.5 \times 10^{20}~\cm^{-2}~ {\rm K}^{-1}~\km^{-1} \s $ in \cite{Ferriere:07} and  $X = 2.8 \times 10^{20}~\cm^{-2}~{\rm K}^{-1}~\km^{-1} \s $ in \cite{Bronfman:88}. Even by taking into account a possible  increasing of $X$ with $r$ (see e.g. eq.(7) in \cite{Ferriere:07}), these values are too different to be compatible. We assume that for $2 < r < 10 $  $X = 1.2 \times 10^{20}~\cm^{-2}~{\rm K}^{-1}~\km^{-1} \s $ as this is the mean value which one gets in that region by taking  $X = 0.5$ at $r = 0$ and assuming that it grows like $\exp(r/7.1)$ as argued in \cite{Arimoto:96}. Hence, we correct Bronfman's et al. density by multiplying it by the factor $(1.2/2.8)$.
\end{description}

\section{Numerical solution of the diffusion equation}
\label{sec:appendix_num}

We will derive here equation (\ref{eq:diff2}).

The diffusion equation which describes the propagation of CRs in a turbulent medium is \eref{j},
\begin{equation}
\nabla_i  J_i (E,r,z) \equiv - \nabla_i\left( D_{ij} (r,z) \nabla_j N(E,r,z)
\right)  = Q (E,r,z)\,.
\end{equation}
Cylindrical symmetry will be assumed.

The diffusion tensor can be conveniently decomposed into
\begin{equation}
D_{ij} = (D_\bot - D_\parallel) b_i b_j  +  D_\parallel \delta_{ij} + D_A \epsilon_{ijk} b_k 
\end{equation}
where $b_i$ are  the components of the regular magnetic field versor.
The symmetric components  $D_\parallel$ and  $D_\bot$  are the diffusion coefficients along 
and  perpendicularly to  ${\bf B}_{\rm reg}$,  while $D_A$  is the antisymmetric 
(Hall) diffusion coefficient. All those coefficients are functions of the energy though with different behaviours. For a fixed value of $L_{\rm max}$  and for $r_L \ll L_{\rm max}$,  $D_\parallel$ and $D_\bot$ are  proportional to $E^{2 - \gamma}$ while $D_A \propto E$  (see e.g. \cite{Ptuskin:93,Candia:04}), where $\gamma$ is the power-law index of the GMF turbulent fluctuations as defined in \sref{subsec:mf}. 
The Hall diffusion becomes dominant only for very high values of the rigidity, i.e. for $r_L(E) \simgeq 0.1 L_{\rm max}$.

As motivated in \sref{subsec:mf} and similarly to what done in other works
\cite{Ptuskin:93,Strong:98,Candia:02}, we can approximate the regular component of the magnetic field to be azimutally oriented, {\rm i.e.}  $b_\phi = \pm 1$ and $b_r = b_z = 0$.
Under such an assumption $D_\parallel$ becomes not physically relevant and the diffusion equation takes the simpler form \cite{Ptuskin:93}
 \begin{equation}
\left\{ -\frac{1}{r} \partial_r \left[ r D_\perp \partial_r \right] - \partial_z \left[ D_\perp \partial_z \right] + u_r \partial_r + u_z \partial_z \right\} N (E,r,z)  = Q (E,r,z)\;,
\end{equation}
where we defined
\begin{eqnarray}
u_r &  \equiv - \partial_z D_A \label{uzeq}  \\
u_z &  \equiv \frac{1}{r} \partial_r \left( D_A r \right)~. \nonumber \label{ureq}
\end{eqnarray}

\end{document}